\documentclass[reprint,aps,prl,10pt,superscriptaddress,amsmath,amssymb,floatfix]{revtex4-2}

\usepackage{amsmath,amssymb,bm}
\usepackage{mathtools}
\usepackage{esvect}
\usepackage{booktabs}
\usepackage{multirow}
\usepackage{array}
\usepackage{upgreek}
\usepackage[T1]{fontenc}
\usepackage[svgnames]{xcolor}
\usepackage[caption=false]{subfig}

\begin{document}

\title{Low-lying $D$ states in yttrium and actinium ions highly sensitive to variation of the fine structure constant}
\author{Akio Kawasaki}
\email{akio.kawasaki@aist.go.jp}
\affiliation{National Metrology Institute of Japan (NMIJ), National Institute of Advanced Industrial Science and Technology (AIST), 1-1-1 Umezono, Tsukuba, Ibaraki 305-8563, Japan}


\begin{abstract}
Whether fundamental constants vary over time or space is one of the key questions in metrology and cosmology. Among them, variation of the fine structure constant $\alpha$ is intensively investigated. Yttrium ions Y$^+$ and actinium ions Ac$^+$ have low-lying $D$ states that are suitable for this search, with a proper path for laser cooling and detection. Theoretical calculations show that the sensitivities of the transitions between the ground state and the lowest $^3D_1$ states are $K=9.40$ and $K=9.73$, respectively. By driving the transition between the ground state and the $^3D_1$ states with a two-photon transition, the transition can be used for a high-sensitivity search for time variation of the fine structure constant. The high efficiency of a detection scheme using the transition between the $7s6d~^3D_1$ states and the $7s7p~^3P_0$ state also suggests that Ac$^+$ ions are potentially useful as a platform for quantum information processing. 

\end{abstract}

\maketitle

Constancy of fundamental constants is one of the fundamental bases of contemporary physics. Violation of this indicates certain kinds of new physics, such as extra dimensions and existence of new particles. This has been intensively investigated using various methods ranging from astronomical observations to experiments in laboratories \cite{RevModPhys.90.025008,ApplPhysRev.12.041331,LivingRevRelativ.28.6}. Particularly, searches for variation of the fine structure constant $\alpha$ have attracted much attention because of potential variation observed in astronomical observations \cite{PhysRevLett.82.884,PhysRevLett.87.091301,SciAdv.6.eaay9672}. However, recent tests using comparisons of different types of atomic clocks rejected certain ranges of variation implied by the astronomical observations. Comparisons of frequencies of two clock transitions in an ytterbium ion Yb$^+$ have put stringent constraints \cite{PhysRevLett.130.253001,PhysRevLett.126.011102,PhysRevLett.113.210801,PhysRevLett.113.210802}. To investigate even fainter variation, the use of narrow-linewidth transitions that has higher sensitivity to variation of $\alpha$ is expected.

Sensitivity to variation of $\alpha$ of a transition is characterized by the following equation \cite{RevModPhys.90.025008,ApplPhysRev.12.041331}. 
\begin{equation}\label{EqAlphaVariation}
\frac{\Delta \nu_0}{\nu_0}=\frac{2q}{h \nu_0}\frac{\Delta \alpha}{\alpha}=K \frac{\Delta \alpha}{\alpha}
\end{equation}
The coefficient $K$ consists of two components: change in the transition frequency $q$ when $\alpha$ varies and energy difference $h\nu_0=hc/k_0$ between the ground state and the excited state, with $h=2\pi \hbar$ being the Planck constant. The former is enhanced when a heavy atom or an orbital with large angular momentum is used, because the sensitivity is a relativistic effect. The sensitivity is enhanced by the latter if the energy difference is small. The transition energies for transitions used for previous searches with optical transitions are around 20000 cm$^{-1}$. If other atoms have excited states with lower energies than these transitions, it is possible that the transition is more sensitive to the variation of $\alpha$, even without going to highly charged ions \cite{RevModPhys.90.045005} or transitions between excited states \cite{PhysRevLett.120.173001,PhysRevA.107.053111}. It should be noted that these excited states has to be metastable to perform precision spectroscopy. 

\begin{figure}[t]
    \subfloat{\includegraphics[width=0.25\textwidth,bb=0 0 500 535]{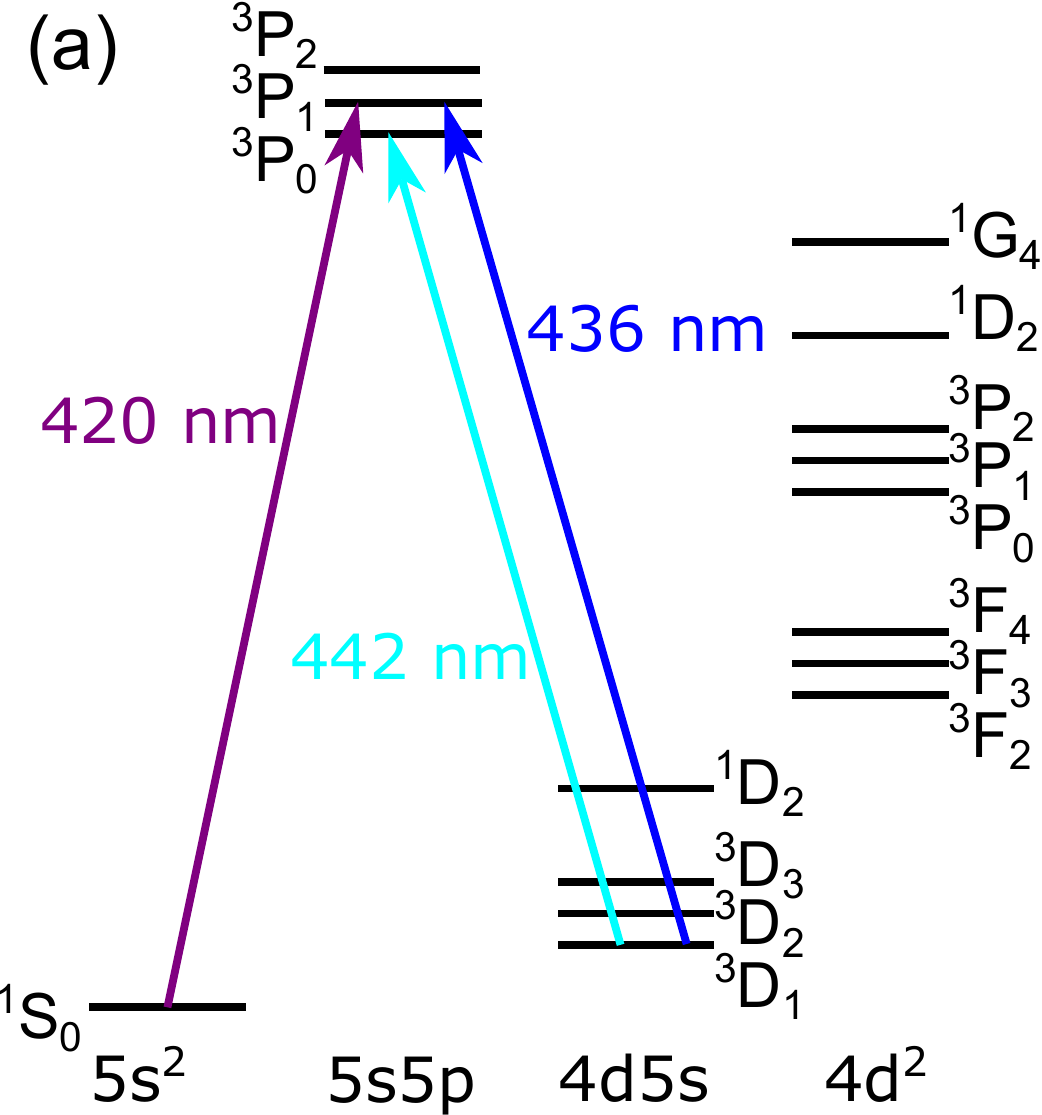}} 
    \subfloat{\includegraphics[width=0.25\textwidth,bb=0 0 500 535]{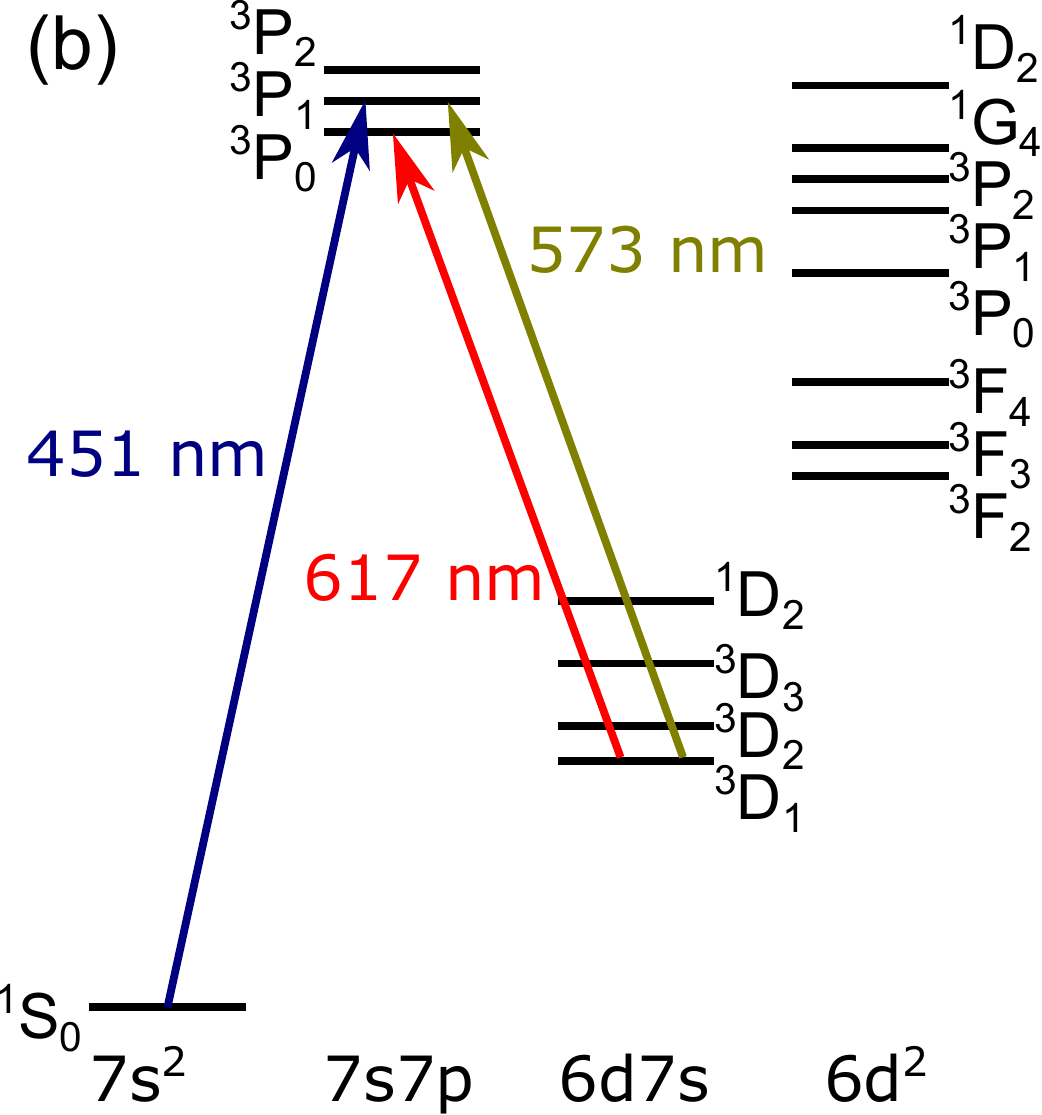}} 
   \caption{Relevant energy levels in (a) Y$^+$ and (b) Ac$^+$. The energy differences are not to scale. }
    \label{FigLevels}
\end{figure}

The survey for such low-energy states is performed using the Atomic Spectra Database at NIST \cite{NISTDB}. The criterion for this survey is to search for low-lying excited states that involve a transition from an $s$ orbital to $d$ or $f$ orbitals. Such traditions can be widely found in transition metals. However, these metals are, in general, not really suitable for laser cooling and efficient detection through closed transitions. In fact, the $5d^{10}6s~^2S_{1/2} \rightarrow 5d^96s^2~^2D_{5/2}$ transition in a gold (Au) atom has $K=8.39$ \cite{PhysRevA.77.012515}. However, Au is not the best atom for laser cooling and detection, which requires five repump lasers. To achieve easy cooling and detection, in this paper, the target of the survey is limited to alkali-atom-like and alkaline-earth-atom-like atoms or singly charged ions. Under this condition, good candidates for the search for the variation of $\alpha$ are the yttrium ion Y$^+$ and the actinium ion Ac$^+$. Relevant energy levels for these two ions are shown in Fig. \ref{FigLevels}. The $D$ states in these two ions have relatively low energies, and particularly for Y$^+$, the energy difference from the ground state is an order of magnitude smaller than the typical energy scale for optical transitions. Therefore, these states are expected to have an enhancement in K due to the small $h\nu_0$ in Eq. \ref{EqAlphaVariation}.

Calculation of the energy levels and $K$ is performed using AMBiT \cite{CompPhysCommun.238.232}. To figure out a proper initial condition for the calculation, different configurations are first tested to optimize the initial condition to minimize the difference in energy levels between the experimental observation and the theoretical calculation for low-lying $S$, $P$, and $D$ states. The result of the calculation for Y$^+$ after the optimization is shown in Table \ref{YResult}. The result shown here is the values after the convergence in a large enough Hilbert space. This result is not as accurate as the one in ref \cite{2604.16274}. Although there were more accurate datasets on the path to the convergence to these numbers, the energies calculated with the largest Hilbert space to which the energy levels converged are chosen as the final result. In general, calculated energy levels are reasonably close to the experimental values with differences within a few percent for most of the states. The reason why the $4d5s$ states appear to have poor relative accuracy is because of the small energy difference between them and the ground state. The absolute inaccuracy is similar to that of other excited states with larger energy. 

Based on this calculation, the sensitivity $K$ to the variation of $\alpha$ for major transitions from the ground state is calculated as shown in Table \ref{YResult}. As expected from the small energy difference from the ground state, the transitions to the $4d5s$ states have relatively large $K$, with the largest value $K=9.40$ for the $4d5s~^3D_1$ state. This number is larger than that for the $6s~^2S_{1/2} \rightarrow 4f^136s^2~2F_{7/2}$ (E3) transition in Yb$^+$ ($K=-5.95$ \cite{CanJPhys.87.25}), which is the transition utilized in the most stringent constraint on $\alpha$ variation \cite{PhysRevLett.130.253001}. 

\begin{table}[tb]
\centering
	\caption{Summary of the theoretical calculation for Y$^+$: Experimental values $E_{\rm exp}$ are cited from Ref. \cite{NISTDB}. Ref. \cite{2604.16274} shows the theoretical values reported in Ref. \cite{2604.16274}. $K$ shows the sensitivity to the variation of $\alpha$ defined in Eq. \ref{EqAlphaVariation}. All energies are shown in units of cm$^{-1}$. }
	\label{YResult}
\begin{tabular}{llrrrr} \toprule
\multicolumn{2}{c}{Configuration}& $E_{\rm exp}$	& Ref. \cite{2604.16274}	& This work	& $K$ \\
 \hline
 \hline	
$5s^2$	& $^1S_0$	& 0			& 0			& 0			& - \\
$4d5s$	& $^3D_1$	& 840.2		& 967.6		& 556.9		& 9.40 \\
$4d5s$	& $^3D_2$	& 1045.1	& 1179.1	& 802.4		& 7.88 \\
$4d5s$	& $^3D_3$	& 1449.8	& 1591.8	& 1155.1	& 6.23 \\
$4d5s$	& $^1D_2$	& 3296.2	& 3429.5	& 2913.6	& 3.24 \\
$5s5p$	& $^3P_0$	& 23445.1	& 23957.5	& 24180.3	& 0.07 \\
$5s5p$	& $^3P_1$	& 23776.2	& 24283.7	& 24511.7	& 0.11 \\
$5s5p$	& $^3P_2$	& 24647.1	& 25167.0	& 25339.7	& 0.26 \\
 \bottomrule
\end{tabular}
\end{table}

The result of the calculation for Ac$^+$ is shown in Table \ref{AcResult}. In the case of Ac$^+$, Ref. \cite{NISTDB} does not have data for Land\'e's $g$ factors, and there have not been any previous reports on theoretical calculations of energy levels. Thus, the theoretical estimate in this report can serve as the first prediction for the $g$ factor. Ref. \cite{NISTDB} has a decent number of the energy levels as experimental values, comparison with which serves as a benchmark for the calculation. The inaccuracy of the energy of the states of our interest is at most a few percent, and the calculation shows reasonable agreement with the experimental values. 

The obtained $K$'s for Ac$^+$ are also large for the transitions to the $6d7s$ states. The largest is $K=9.73$ for the $7s^2~^1S_0 \rightarrow 6d7s~^3D_1$ transition. All numbers for the $D$ states shown here are larger than that for the E3 transition in Yb$^+$. It should also be noted that the sign is opposite to that of the value for the E3 transition in Yb$^+$. By comparing the E3 transition in Yb$^+$ and the $7s^2~^1S_0 \rightarrow 6d7s~^3D_1$ transition in Ac$^+$, the sensitivity can be further enhanced. 
 
\begin{table}[tb]
\centering
	\caption{Summary of the theoretical calculation for Ac$^+$: Experimental values $E_{\rm exp}$ are cited from Ref. \cite{NISTDB}. $g$ factor shows Land\'e's $g$ factor $g_J$. $K$ shows the sensitivity to the variation of $\alpha$ defined in Eq. \ref{EqAlphaVariation}. All energies are shown in units of cm$^{-1}$. }
	\label{AcResult}
\begin{tabular}{llrrrr} \toprule
\multicolumn{2}{c}{Configuration}& $E_{\rm exp}$	& This work	& $g$ factor	& $K$ \\
 \hline
 \hline	
$7s^2$	& $^1S_0$	& 0			& 0			& 0		& - \\
$6d7s$	& $^3D_1$	& 4739.6	& 4598.3	& 0.499	& 9.73 \\
$6d7s$	& $^3D_2$	& 5267.1	& 5222.7	& 1.147	& 9.01 \\
$6d7s$	& $^3D_3$	& 7426.5	& 7252.2	& 1.334	& 6.57 \\
$6d7s$	& $^1D_2$	& 9087.5	& 8896.3	& 1.007	& 6.94 \\
$7s7p$	& $^3P_0$	& 20956.4	& 21733.5	& 0		& 0.37 \\
$7s7p$	& $^3P_1$	& 22180.5	& 22915.9	& 1.407	& 0.55 \\
$7s7p$	& $^3P_2$	& 28201.1	& 27665.1	& 1.488	& 1.06 \\
 \bottomrule
\end{tabular}
\end{table}

To make full use of these highly sensitive transitions for searches for variation of $\alpha$, a proper clock operation is necessary. As for the  $7s^2~^1S_0 \rightarrow 6d7s~^3D_1$ transition in Ac$^+$, the wavelength of 2110 nm is within the range of commercially available lasers. This means that, in principle, we can drive the transition directly, though it would suffer from a large probe-induced AC Stark shift. However, for the $5s^2~^1S_0 \rightarrow 4d5s~^3D_1$ transition in Y$^+$, the frequency of 25.19 THz is not in a typical range for a narrow-linewidth laser. This makes it difficult for the transition to be directly driven by a single frequency light. Instead, a two-photon transition through the $5s5p~^3P_1$ state can be implemented using two lasers at 420 nm and 436 nm, as shown in Fig. \ref{FigLevels}(a). A similar two-photon transition can be achieved for Ac$^+$ as well through the $7s7p~^3P_1$ state, using 451 nm and 573 nm lasers. Such a two-photon transition for clock operation is widely used for coherent population trapping (CPT) clocks with cesium or rubidium \cite{ApplPhysB.81.421}. In CPT clocks, a single laser is typically modulated by a radio-frequency (RF) source, whereas in the $5s^2~^1S_0 \rightarrow 4d5s~^3D_1$ transition in Y$^+$ has too large an energy difference for such modulation by an RF source. Instead, two lasers with different frequencies should be used, with their frequency difference fixed by phase-locking to a frequency comb. Such an implementation of a two-photon transition with two lasers of completely different frequencies is realized in the field of molecular quantum gases \cite{NewJPhys.17.075016}. 

To estimate systematic uncertainties for this two-photon transition operation, matrix elements for major transitions in Y$^+$ and Ac$^+$ are calculated with the AMBiT. The result for Y$^+$ is shown in Table \ref{YTransitionRate} together with the transition rates listed in Ref. \cite{NISTDB}. Most of the transitions have a reasonable agreement between the theoretical calculation and the experimental data, with a couple of transition rates being off by an order of magnitude. Because all of these data with large discrepancies involve the $5s5p~^3P_2$ state, there can be some large inaccuracy in the calculation for the $5s5p~^3P_2$ state. Nevertheless, the overall lifetime estimate has a reasonable agreement with another theoretical estimate in Ref. \cite{2604.16274}, as shown in Table \ref{YLifetime}. These comparisons show that the matrix element calculation is reliable, at least to the extent of providing a rough estimate of transition rates. The same comparison between the theoretical calculation and experimental data is performed for Ac$^+$ as well, as shown in Table \ref{AcTransitionRate}. The Ac$^+$ calculation has better agreement between the theory and the experiment than the Y$^+$ calculation. The lifetime of the $6d7s~^3D_1$ state is estimated to be $3.0\times10^5$ s, which is long enough to serve as as a metastable state suitable for precision spectroscopy. 

\begin{table}[tb]
\centering
	\caption{Transition rates $A_{ki}$ of major transitions in Y$^+$: $|i\rangle$ and $|k\rangle$ are the low-energy and high-energy states addressed by the transition. $S_{ik}$ is the matrix element calculated by AMBiT in this work. It is shown in atomic units. $\lambda_{ik}$ shows the wavelength of the light resonant to the transition displayed in the units of nm. Transition rate $A_{ki}$ is shown in units s$^{-1}$. Exp. shows the experimental values listed in Ref. \cite{NISTDB}.}
	\label{YTransitionRate}
\begin{tabular}{llllrrrr} 
\toprule
\multicolumn{2}{c}{$|i\rangle$}& \multicolumn{2}{c}{$|k\rangle$} 	& $\lambda_{ik}$	& $S_{ik}$	& \multicolumn{2}{c}{$A_{ki}$}  \\
 		& 			& 			&  			& 			&		& This work			& Exp.  \\
 \hline
 \hline	
$4d5s$	& $^3D_1$	& $5s5p$	& $^3P_0$	& 442.38	& 0.859	& $2.01\times 10^7$	& $1.83\times 10^7$	\\
$5s^2$	& $^1S_0$	& $5s5p$	& $^3P_1$	& 420.59	& 0.343	& $3.11\times 10^6$	& $2.19\times 10^6$	\\
$4d5s$	& $^3D_1$	& $5s5p$	& $^3P_1$	& 435.99	& 0.779	& $6.35\times 10^6$	& $5.55\times 10^6$	\\
$4d5s$	& $^3D_2$	& $5s5p$	& $^3P_1$	& 439.92	& 1.446	& $1.15\times 10^7$	& $1.151\times 10^7$	\\
$4d5s$	& $^1D_2$	& $5s5p$	& $^3P_1$	& 488.28	& 0.0108	& $6.26\times 10^4$	& $1.50\times 10^5$	\\
$4d5s$	& $^3D_1$	& $5s5p$	& $^3P_2$	& 420.05	& 0.749	& $4.09\times 10^6$	& $5.35\times 10^5$	\\
$4d5s$	& $^3D_2$	& $5s5p$	& $^3P_2$	& 423.69	& 0.0341	& $1.82\times 10^5$	& $2.32\times 10^6$	\\
$4d5s$	& $^3D_3$	& $5s5p$	& $^3P_2$	& 431.08	& 1.741	& $8.81\times 10^6$	& $1.285\times 10^7$	\\
$4d5s$	& $^1D_2$	& $5s5p$	& $^3P_2$	& 468.36	& 7.444	& $2.94\times 10^7$	& $1.88\times 10^6$	\\
 \bottomrule
\end{tabular}
\end{table}

\begin{table}[tb]
\centering
	\caption{Lifetime of major states in Y$^+$ in units of s.}
	\label{YLifetime}
\begin{tabular}{llrr} 
\toprule
\multicolumn{2}{c}{Configuration} & This work	& Ref. \cite{2604.16274} \\
 \hline
 \hline	
$5s5p$	& $^3P_0$	& $4.8\times 10^{-8}$	& $5.4(4)\times 10^{-8}$ \\
$5s5p$	& $^3P_1$	& $4.5\times 10^{-8}$	& $5.1(4)\times 10^{-8}$ \\
$5s5p$	& $^3P_2$	& $2.3\times 10^{-8}$	& $5.6(1.0)\times 10^{-8}$ \\
$4d5s$	& $^3D_1$	& $5.9\times 10^{10}$	& $4.5(5)\times 10^{10}$ \\
 \bottomrule
\end{tabular}
\end{table}

\begin{table}[tb]
\centering
	\caption{Transition rates $A_{ki}$ of major transitions in Ac$^+$: $|i\rangle$ and $|k\rangle$ are the low-energy and high-energy states addressed by the transition. $S_{ik}$ is the matrix element calculated by AMBiT in this work, shown in atomic units. $\lambda_{ik}$ shows the wavelength of the light resonant to the transition displayed in units of nm. Transition rate $A_{ki}$ is shown in units of s$^{-1}$. Exp. shows the experimental values listed in Ref. \cite{NISTDB}. }
	\label{AcTransitionRate}
\begin{tabular}{llllrrrr} 
\toprule
\multicolumn{2}{c}{$|i\rangle$}& \multicolumn{2}{c}{$|k\rangle$} 	& $\lambda_{ik}$	& $S_{ik}$	& \multicolumn{2}{c}{$A_{ik}$}  \\
 		& 			& 			&  			& 			&		& This work			& Exp.  \\
 \hline
 \hline	
$6d7s$	& $^3D_1$	& $7s7p$	& $^3P_0$	& 616.30	& 3.237	& $2.80\times 10^7$	& $2.80\times 10^7$	\\
$7s^2$	& $^1S_0$	& $7s7p$	& $^3P_1$	& 450.85	& 2.655	& $1.96\times 10^7$	& $2.13\times 10^7$	\\
$6d7s$	& $^3D_1$	& $7s7p$	& $^3P_1$	& 573.36	& 2.591	& $9.28\times 10^6$	& $1.04\times 10^7$	\\
$6d7s$	& $^3D_1$	& $7s7p$	& $^3P_2$	& 426.23	& 0.0181	& $9.46\times 10^4$	& $1.70\times 10^5$	\\
 \bottomrule
\end{tabular}
\end{table}

\begin{figure*}[t]
    \subfloat{\includegraphics[width=0.45\textwidth, bb=0 0 699 436]{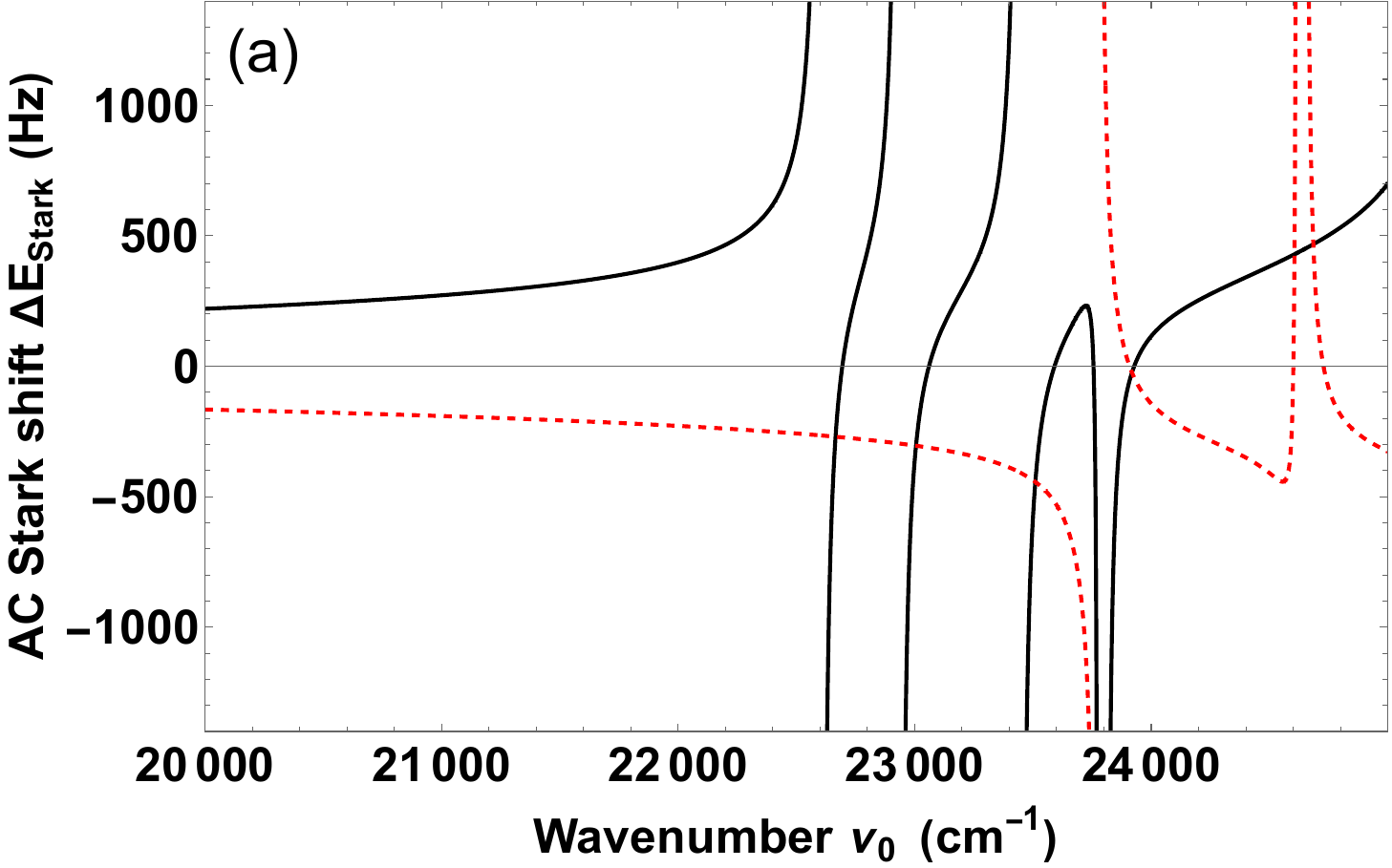}} 
\hspace{0.04\textwidth}
    \subfloat{\includegraphics[width=0.45\textwidth, bb= 0 0 697 435]{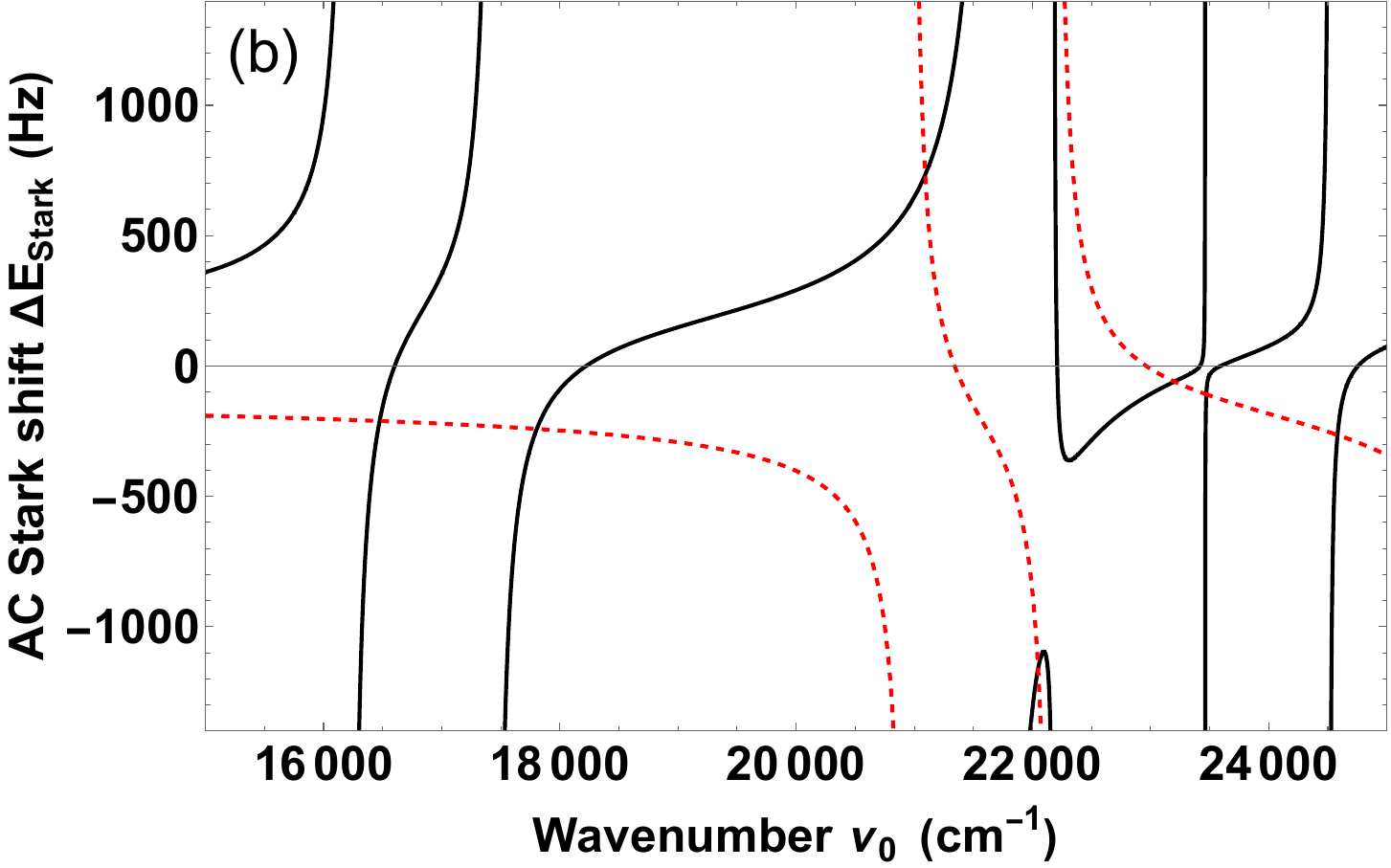}} 
   \caption{AC Stark shifts induced by light of two frequencies that induce the two-photon transition between the ground state and the low-lying $^3D_1$ state for (a) Y$^+$ and (b) Ac$^+$: The black solid line shows the AC Stark shift induced by the laser with the higher frequency. The red dashed line shows the AC Stark shift induced by the laser with the lower frequency with its sign flipped so that the crossing point can show the laser frequency where the total AC Stark shift is cancelled between these two lasers. The horizontal axis shows the wavenumber of a photon for the higher frequency. The energy for the lower-frequency laser is set so that the frequency difference corresponds to the energy difference between the ground state and the $^3D_1$ state. The power for each laser is assumed to be an intensity of 1.33 mW/cm$^2$. }
   \label{FigPolarizability} 
\end{figure*}

The main benefit of the two-photon operation for the clock is the cancellation of the AC Stark shift induced by the probe laser. To demonstrate this, the AC Stark shift induced by two lasers for the two-photon transition is calculated. For simplicity, the following discussion is limited to the leading-order scalar shift induced by electric dipole (E1) transitions. It should be noted that vector and tensor shifts can be cancelled by properly chosen polarization and magnetic field configurations. This AC Stark shift $\Delta E_{\rm Stark}=-\frac{1}{4}\alpha_{n} {\cal E}^2$ \cite{PhysRevLett.91.173005} of a state $n$ is characterized by the polarizability $\alpha_n$ by the following equation: 
\begin{equation}
\alpha_n=\frac{2}{\hbar}\sum_k \frac{\omega_{kn}S_{kn}}{\omega_{kn}^2-\omega^2}, 
\end{equation}
where $S_{kn}$ and $\omega_{kn}$ are the matrix elements for the E1 transition and the resonant frequency for the transition between state $n$ and $k$, respectively. Because $\alpha_n$ changes its sign at the resonance, the near-resonant behavior is important to see the overall behavior, whereas far off-resonant light adds an offset almost constant over $\omega$. 

For Y$^+$, the five lowest-energy states for each combination of $J$ and parity allowed for the E1 transitions are the dominant contributors to the AC Stark shift. Because corresponding states in Ref. \cite{NISTDB} are identified for these states in the theoretical calculation, the energy difference used for the AC Stark shift calculation is based on Ref. \cite{NISTDB}, whereas the matrix elements are based on the theoretical calculation. The resulting AC Stark shift is shown in Fig. \ref{FigPolarizability}(a). The behavior of the AC Stark shift is basically that of multiple resonance features. Thus, essentially, the AC Stark shift induced by one laser goes from negative infinity to positive infinity repeatedly, while the other laser does not necessarily hit a resonance. It is easy to find zero-crossings, which are suitable for the spectroscopy free from the AC Stark shift. Among them, the one at 23918.6 cm$^{-1}$ has a desirable behavior for precision spectroscopy: small shifts for both lasers are cancelled out. This is 4.268 THz away from the resonance from the ground state to the $5s5p~^3P_1$ state. Note that this is 3.349 THz away from the $4d5s~^3D_1 \rightarrow 5s5p~^3P_2$ transition, which can induce an erroneous excitation to the $5s5p~^3P_2$ state if the initial state is the $4d5s~^3D_1$ state. This is because the energy difference between the $5s5p~^3P_2$ and the $5s5p~^3P_1$ states is coincidentally close to that between the ground state and the $4d5s~^3D_1$ state, and such potential excitation to a wrong state is unavoidable. The two photon Rabi frequency at this frequency is $2\pi \times 19.8$ Hz for an intensity of 133 mW/cm$^2$ for both lasers, which is reasonable for the standard atomic clock operation. 

For Ac$^+$, the six lowest-energy states for each combination of $J$ and parity are included in the calculation. Energy levels based on the theoretical calculation are used only if the state is not identified in Ref. \cite{NISTDB}. The result of the AC Stark calculation is shown in Fig. \ref{FigPolarizability}(b). Here, a convenient zero crossing is the one at 23201.41 cm$^{-1}$, which is far from any strong transition and has a relatively small shift for both lasers. This is 30.60 THz away from the resonance from the ground state to the $7s7p~^3P_1$ state. The Rabi frequency for this transition is $2\pi \times 14.0$ Hz for an intensity of 133 mW/cm$^2$ for the two lasers, which is also reasonably high for the standard atomic clock operation. It should be noted that the zero crossing can be engineered by adjusting the power ratio of two lasers. 

Another advantage of the two-photon transition is the cancellation of the first-order Doppler shift, which can be achieved by counterpropagating two lasers that are used for the interrogation. This suppresses the systematic uncertainties to some extent \cite{PhysRevLett.123.033201}, but because the major source of the uncertainty due to the Doppler shift is not the first order but the second order, the benefit is somewhat minor. 

\begin{table}[b]
\centering
	\caption{Decay rates of the $7s7p~^3P_0$ transition in Ac$^+$: Conf. shows the electronic configuration of the final state of the decay. E1, E2, M1, and M2 in the Tr. column show allowed decay channels, corresponding to the electric dipole, electric quadrupole, magnetic dipole, and magnetic quadrupole transitions, respectively. $S_{ik}$ is shown in atomic units. Please note that the decay to $J=0,3,$ and $4$ states are not allowed based on the selection rules for E1, E2, M1, and M2 transitions.}
	\label{Ac3P0DecayRate}
\begin{tabular}{llrrrrr} 
\toprule
\multicolumn{2}{c}{Conf.}	& Tr. 	& $\lambda_{ik}$ (nm)	& $S_{ik}$	& $A_{ik}$ (s$^{-1}$)	& Ratio \\
\\
 \hline
 \hline	
$6d7s$	& $^3D_1$	& E1	& 616.30	& 3.237		& $2.80\times 10^7$	& 0.99993	\\
$6d^2$	& $^3P_1$	& E1	& 5127.84	& 0.131		& $1.97\times 10^3$	& $7.02\times 10^{-5}$	\\
$6d7s$	& $^3D_2$	& M2	& 637.01	& 0.02208	& $3.14\times 10^{-8}$	& $1.12\times 10^{-15}$	\\
$6d7s$	& $^1D_2$	& M2	& 841.90	& 35.54		& $1.25\times 10^{-5}$	& $4.47\times 10^{-13}$	\\
$6d^2$	& $^3F_2$	& M2	& 1293.83	& 2.671		& $1.10\times 10^{-7}$	& $3.92\times 10^{-15}$	\\
$6d^2$	& $^3P_2$	& M2	& 5673.89	& 10.70		& $2.71\times 10^{-10}$	& $9.68\times 10^{-18}$	\\
 
\bottomrule
\end{tabular}
\end{table}

The disadvantage of the clock with low-lying $^3D_1$ states is a large blackbody radiation (BBR) shift. This is characterized by the differential DC polarizability, which is 79.64 kHz/(kV/cm)$^2$ and 145.20 kHz/(kV/cm)$^2$ for Y$^+$ and Ac$^+$, respectively. This is larger than Sr, which is known to have a relatively large DC polarizability of $61.5558(17)$ kHz/(kV/cm)$^2$ \cite{PhysRevLett.109.263004}. To accurately operate Y$^+$ and Ac$^+$ clocks, a cryogenic system might be necessary, not only to stabilize the temperature of the vacuum chamber. This large polarizability at the low-frequency limit induces a large probe AC Stark shift, if the transition to the $^3D_1$ states are driven directly by a light field. The differential polarizability at the resonance of the transition from the ground state to the $^3D_1$ state is 79.73 and 151.80 kHz/(kV/cm)$^2$ for Y$^+$ and Ac$^+$, respectively. 

To operate atomic clocks, a proper detection scheme is necessary. For Y$^+$, Ref. \cite{2604.16274} claims that the $4d5s~^3D_1 \rightarrow 5s5p~^3P_0$ transition is nearly closed and good for detection. To test whether the $6d7s~^3D_1 \rightarrow 7s7p~^3P_0$ transition in Ac$^+$ can serve as a closed transition for detection in a similar way, relevant transition rates are calculated, as shown in Table \ref{Ac3P0DecayRate}. This shows that the $6d7s~^3D_1 \rightarrow 7s7p~^3P_0$ transition is essentially closed with a branching ratio of 99.993 \%. Please note that the branching ratio of 5 \% for the $4d5s~^3D_1 \rightarrow 5s5p~^3P_0$ transition in Ref. \cite{2604.16274} is reproduced as 3.5 \% in the calculation in this work. This means that the $6d7s~^3D_1 \rightarrow 7s7p~^3P_0$ transition is much more suitable for the detection than the $4d5s~^3D_1 \rightarrow 5s5p~^3P_0$ transition. It also means that Ac$^+$ can be a good platform for not only for searches for time variation of $\alpha$ but also for quantum information processing in the same way as Y$^+$. Note that the isotope of Ac easiest to obtain is $^{227}$Ac, which has a lifetime of 21.77 yr and is a part of the actinium series. This has a nuclear spin of $I=3/2$, which results in a complicated hyperfine structure. For the simplest hyperfine structure, $^{224}$Ac has $I=0$. However, its lifetime is 2.78 h and it is somewhat difficult to generate, as the main decay channel to $^{224}$Ac is an $\alpha$ decay of $^{228}$Pa, which has a branch ratio of 1.85\% and a lifetime of 22 h.

To summarize, sensitivity to variation of the fine structure constant is calculated for transitions in Y$^+$ and Ac$^+$. Low-lying $D$ states showed high sensitivity. Particularly, the $5s^2~^1S_0 \rightarrow 4d75~^3D_1$ transition in Y$^+$ and the $7s^2~^1S_0 \rightarrow 6d7s~^3D_1$ transition in Ac$^+$ have high sensitivity of $K=9.40$ and $K=9.73$, respectively. Clock operation with these low-lying excited states can be performed using a two-photon transition, which has an advantage of the cancellation of probe-induced AC Stark shift, whereas the BBR shift is expected to be high for these transitions. Regarding Ac$^+$, the detection scheme to use the $6d7s~^3D_1 \rightarrow 7s7p~^3P_0$ transition is promising because this transition is almost closed, and this potentially opens a way to make use of Ac$^+$ for a new platform for quantum information processing. 

~

{\it Acknowledgments}---This work was supported by Japan Society for the Promotion of Science KAKENHI Grants No. 26H00406 and Japan Science and Technology Agency FOREST Grant No. JPMJFR212S.

~ 

{\it Data Availability}---The data that support the findings of this article are not publicly available. The data are available from the authors upon reasonable request.

\bibliography{YAc}

\end{document}